\documentclass[aps,prl,reprint,groupedaddress]{revtex4-2}

\usepackage{amsmath}
\usepackage{calrsfs}
\usepackage{mathbbol}
\usepackage{longtable}
\usepackage{graphicx}
\usepackage{amssymb}

\begin{document}


\title{Polar phase transition in $\mathbf{180}^{\boldsymbol{\circ}}$-domain wall of lead titanate}



\author{I.~Rychetsky}
\email{rychet@fzu.cz}
\affiliation{Institute of Physics of the Czech Academy of Sciences, Na Slovance 2, 18221 Prague 8, Czech Republic.}
\author{W.~Schranz}
\affiliation{University of Vienna, Faculty of Physics, Boltzmanngasse 5, 1090 Wien, Austria.}
\author{A.~Tr\"{o}ster}
\affiliation{University of Vienna, Faculty of Physics, Boltzmanngasse 5, 1090 Wien, Austria.}

\date{\today}

\begin{abstract}
A new mechanism leading to a switchable polarization in a ferroelectric  domain wall (DW) is proposed. A biquadratic coupling of the primary order parameter and its gradient triggers the phase transition in the DW with softening of the local polar mode and anomalous increase of the  susceptibility at the phase transition temperature $T_{DW}$. This mechanism describes the origin and properties of the polar \textit{Bloch} and antipolar \textit{N\'eel} components in the $180^\circ$-DW of PbTiO$_3$, which were recently reported from first-principles calculations.
\end{abstract}


\maketitle

\textit{Introduction.}---The tensor properties of domain walls (DWs) in ferroic materials become recently of increasing interest driven by achievements in technological and measurement methods allowing to fabricate and observe submicron and nanoscale structures. Various methods for modeling of DWs are widely used \cite{meier2020}, i.~e. first-principle  calculations \cite{Iniguez2020}, machine-learned force fields \cite{Troester2022}, phase-field modeling \cite{Voelker2011}, and phenomenological Landau-Ginzburg theory \cite{Marton2010}, which are closely interconnected with the DW symmetry analysis described by layer groups \cite{Schranz2019,Schranz2020_JAP,Schranz2020_STO,Rychetsky2021_PZO,Schranz2022}. 
Polarization inside DWs was predicted in some perovskite structures \cite{Stepkova2012,Marton2013}, where the crucial role was assigned to flexoelectricity \cite{Eliseev2013}, rotopolar coupling \cite{Stengel2017,Schranz2020_STO} or biquadratic coupling of the primary and secondary order parameters \cite{Tagantsev2001}. 

The possible existence of polar ${180}^{\boldsymbol{\circ}}$-DW in $\mathrm {PbTiO}_3$ (PTO) was reported by several authors. However, the situation is not so clear yet. Based on \textit{ab initio} calculations an \textit{Ising} structure of the DW profile was reported in Ref.~\cite{Meyer2002}. Such a DW would not carry any polarization within the wall. Other authors concluded that the DW contains also a \textit{N\'eel}-like polarization (asymmetric polarization profile) originating from flexoelectricity \cite{Wang2017,Behera2011} and a switchable \textit{Bloch} component indicating a ferroelectric phase transition inside the DW \cite{Wojdel2014,Iniguez2020}. The latter behavior was not found to be stable within the Landau-Ginzburg approach \cite{Behera2011,Wang2017}, where only the \textit{N\'eel} polarization was obtained.
In this contribution we show that the symmetry of the DW (layer group) together with an extended Landau-Ginzburg potential allow to  properly describe the polar properties of ${180}^{\boldsymbol{\circ}}$-DW in PTO.
  
\textit{Symmetry of $180^{\boldsymbol{\circ}}$DW}---
PTO exhibits a uniaxial ferroelectric phase transition from cubic to tetragonal structure without multiplication of the unit cell. The symmetry decrease from $Pm\bar{3}m$ to $P4mm$ implies 6 tetragonal domain states (DSs) $1_{1}\equiv (-P_s,0,0)$, $2_1\equiv (0,-P_s,0)$, $3_1\equiv (0,0,-P_s)$ and $1_2,\ 2_2,\ 3_2$ with opposite sign of polarization. 

Here we consider the $\mathbf{180}^{\boldsymbol{\circ}}$--DW $(3_1|\mathbf{n,p}|3_2)$ between the DSs $3_1\equiv (0,0,-P_s)$ and $3_2\equiv (0,0,P_s)$, with the normal $\mathbf{n}\parallel x$ and the microscopic position within the unit cell $\mathbf{p}$ \cite{Schranz2020_STO,Rychetsky2021_PZO}. The macroscopic tensor properties of DWs described by Landau theory are independent of the microscopic position $\mathbf{p}$ and they are determined by the layer group symmetry of the DW twin $(3_1|\mathbf{n}|3_2)$, which contains 4 elements $T_{12}=\mathbf{T}\{1,m_y,2_y,\bar{1}\}$, $\mathbf{T}$ are translations parallel with the DW plane \cite{Rychetsky2021_PZO}. This symmetry implies that the \textit{N\'eel} component is antisymmetric, $P_1(x)=-P_1(-x)$, and it can be nonzero in the whole temperature range below $T_c$. The \textit{Bloch} component is forbidden by symmetry, since application of $m_y$ yields $P_2(x)=-P_2(x)=0$. Therefore it could only occur as a result of the phase transition lowering the symmetry to $T'_{12}=\mathbf{T}\{1,2_y\}$. Then the \textit{Bloch} component is nonzero and symmetric: $P_2(x)=P_2(-x)\neq 0$. The polarization profiles and the phase transition in the DW are further analyzed using the Landau-Ginzburg free energy description. 

\textit{The free energy}---
The Gibbs free energy can be written as:
\begin{equation}
	G(\mathbf{P,\boldsymbol{\sigma}})=G_0+G_{es}+G_{el}+G_{flex}+G_{biq}+G_{g}
\end{equation}
where the individual parts, pure polarization $G_0$, electrostriction $G_{es}$, elastic energy $G_{el}$, gradient term $G_g$, flexoelectric $G_{flex}$, biquadratic OP and its gradient $G_{biq}$ read
\begin{widetext}
\begin{eqnarray}
	G_0&=& \alpha_{1} \left(P_1^2+P_2^2+P_3^2\right)+ 
	\alpha_{11}\left(P_1^4+P_2^4+P_3^4\right)+ \alpha_{12} \left(P_1^2
	P_2^2+P_3^2 P_2^2+P_1^2 P_3^2\right) \nonumber\\
	&+& \alpha_{123} P_1^2 P_2^2
	P_3^2 +\alpha_{111}\left(P_1^6+P_2^6+P_3^6\right)
	+     \alpha_{112}\left(\left(P_2^4+P_3^4\right) P_1^2+\left(P_1^4+P_2^4\right)
	P_3^2+P_2^2 \left(P_1^4+P_3^4\right)\right)\\
	G_{es}&=&-\sigma_1 \left(P_1^2  Q_{11}+P_2^2  Q_{12}+P_3^2
	Q_{12}\right)-\sigma_2 \left(P_2^2  Q_{11}+P_1^2
	Q_{12}+P_3^2  Q_{12}\right)
	-\sigma_3 \left(P_3^2
	Q_{11}+P_1^2  Q_{12}+P_2^2  Q_{12}\right) \nonumber\\
	&-& Q_{44} (P_2
	P_3 \sigma_4+P_1 P_3 \sigma_5+P_1 P_2 \sigma_6)\\
	G_{el}&=&-\frac{1}{2}
	\left(\left(\sigma_1^2+\sigma_2^2+\sigma_3^2\right)  s_{11}+2
	(\sigma_1 \sigma_2+\sigma_3 \sigma_2+\sigma_1 \sigma_3)
	s_{12}+\left(\sigma_4^2+\sigma_5^2+\sigma_6^2\right)
	s_{44}\right)\\
	G_g&=&\frac{1}{2}\left(g_{11}\left(\frac{\partial P_1}{\partial x}\right)^2+g_{44}\left(\frac{\partial P_2}{\partial x}\right)^2+g_{44}\left(\frac{\partial P_3}{\partial x}\right)^2\right) \\
	G_{flex}&=&   
	- \frac{\partial P_1}{\partial x}\left(f'_{14}\left(P_{2}^2 
	+ P_{3}^2\right) 
	+f_{11} \sigma_{1}
	+ f_{14}\left(\sigma_{2} 
	+ \sigma_{3}\right)
	\right) 
	- f_{111}\left(\frac{\partial P_2}{\partial x}\sigma_{6} 
	+ \frac{\partial P_3}{\partial x}\sigma_{5}\right) \\
	G_{biq}&=&f_{22}\left[ 
	P_2^2\left(\frac{\partial P_3}{\partial x}\right)^2 +
	P_3^2\left(\frac{\partial P_2}{\partial x}\right)^2
	\right] +f'_{22}\left[ 
	\sigma_2\left(\frac{\partial P_3}{\partial x}\right)^2 +
	\sigma_3\left(\frac{\partial P_2}{\partial x}\right)^2
	\right]
\end{eqnarray}
\end{widetext}
$G_{flex}$ and $G_{biq}$ are terms not considered in \cite{Behera2011}. Since the DW properties are $x$-dependent the gradient terms contain only $\partial\square/\partial x$ derivatives. 
The quasi-1D DW along $x$-axis requires mechanical equilibrium  $\sigma_{1}=\sigma_{5} =\sigma_{6}=0$ and compatibility of strains $e_2(x)=e_{2s},e_3(x)=e_{3s},e_4(x)=e_{4s}$, where $e_{is}$ are spontaneous strains of homogeneous domains. Therefore it is convenient to use the thermodynamic potential $F(\mathbf{P},\sigma_1,\sigma_2,\sigma_6,e_2,e_3,e_4)$ obtained by the Legendre transformation:
$F=G+\sigma_2 e_2+\sigma_3 e_3+\sigma_4 e_4$. For the sake of simplicity it is also convenient to assume $f_{14}=f'_{22}=0$, since it only renormalizes some coefficients but does not change the overall polarization behavior. Taking into account all above the potential $F$ is expressed as: 
\begin{equation}\label{eq:f}
	F=F_0+F_{flex}+F_{biq},
\end{equation}
where
\begin{widetext}
	\begin{eqnarray}
		F_0&=& b_{1} P_1^2+b_2 P_2^2+b_3 P_3^2+ 
		b_{11}P_1^4+b_{22}(P_2^4+P_3^4)+ b_{12} (P_1^2
		P_2^2+P_1^2 P_3^2) + b_{23}P_2^2 P_3^2 \nonumber\\
		&+& \alpha_{123} P_1^2 P_2^2
		P_3^2 +\alpha_{111}\left(P_1^6+P_2^6+P_3^6\right)
		+     \alpha_{112}\left(\left(P_2^4+P_3^4\right) P_1^2+\left(P_1^4+P_2^4\right)
		P_3^2+P_2^2 \left(P_1^4+P_3^4\right)\right) \nonumber\\
		&+&\frac{1}{2}\left(g_{11}\left(\frac{\partial P_1}{\partial x}\right)^2+g_{44}\left(\frac{\partial P_2}{\partial x}\right)^2+g_{44}\left(\frac{\partial P_3}{\partial x}\right)^2\right) \label{eq:f0}\\
		F_{flex}&=&   
		- f'_{14}\frac{\partial P_1}{\partial x}\left(P_{2}^2 
		+ P_{3}^2\right) 
		\label{eq:fflex}\\
		F_{biq}&=&f_{22}\left[ 
		P_2^2\left(\frac{\partial P_3}{\partial x}\right)^2 +
		P_3^2\left(\frac{\partial P_2}{\partial x}\right)^2
		\right] 
		\label{eq:fbiq}
		\end{eqnarray}
\end{widetext}
The value of spontaneous polarization is $
P_s=\sqrt{\frac{-a_{11}+\sqrt{a_{11}^2-3 a_1 a_{111}}}{3 a_{111}}}
$. The $b-$coefficients are explicitly written in \textit{Appendix~A}, and the numerical values of coefficients for PTO are shown in TABLE~\ref{tab:1}. Since for further considerations $f_{22}\le 0$, the stability condition requires $f_{22}P_s^2+g_{44}/2>0$. 
$F_0$ was already discussed in \cite{Behera2011}. It is shown below that $F_0$ alone does not lead to the DW polarization, while the flexoelectric coupling induces the \textit{N\'eel} component $P_1$, and the biquadratic coupling of the OP and its gradient can cause the appearance of the \textit{Bloch} component $P_2$.
\begin{figure}
	\includegraphics[width=\linewidth]{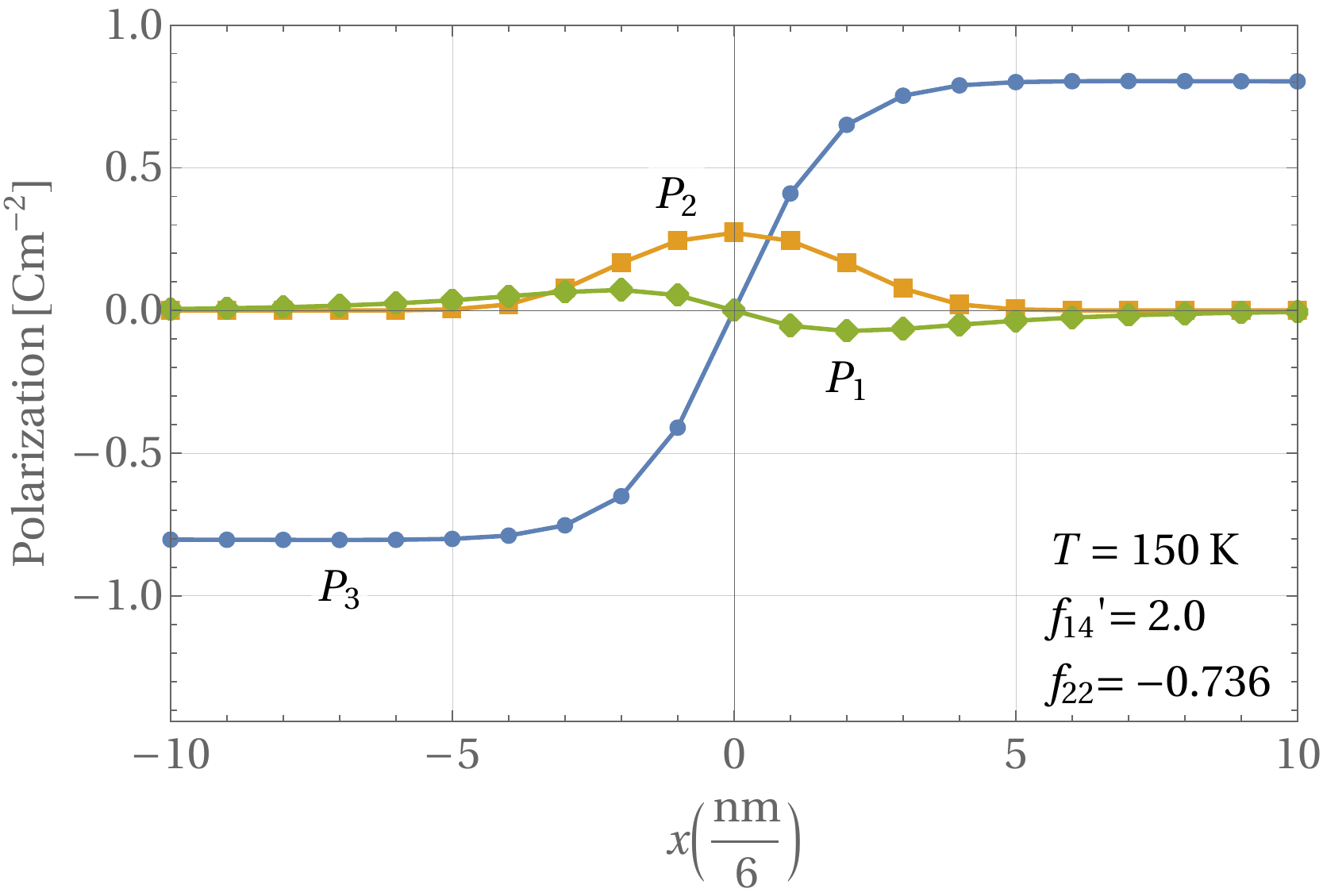}
	\caption{\label{fig:profile}
		The most general profile, mixed \textit{Bloch}+\textit{N\'eel}, at low temperatures, $P_1\ne0,\ P_2\ne0$. Qualitatively, in the \textit{Bloch} profile $P_1=0,\ P_2\ne0$, in the \textit{N\'eel} profile $P_1\ne0,\ P_2=0$, and in the \textit{Ising} profile $P_1=0,\ P_2=0$.}
\end{figure}
\\
\textit{$180^{\circ}$DW}---The polarization profile can be obtained by minimizing the free energy functional\\ $
\mathcal{L}=\int_{-\infty}^{\infty}F(\mathbf{P}(x),\partial_x\mathbf{P}(x))dx
$ with proper boundary conditions. In practice, this can be achieved by direct minimization of the discretized (finite difference) free energy. An example of the DW profile at low temperatures is shown in Fig.~\ref{fig:profile}. Alternatively, if possible, it can be obtained by solving Lagrange-Euler (LE) equations. Let us first assume $F=F_0$ , i.~e. $f'_{14}=f_{22}=0$. Then the LE equations can be solved explicitly and the \textit{Ising} DW profile is obtained:
\begin{equation}\label{eq:ising}
	P_1=P_2=0,\ P_3=\frac{P_s\tanh(x/2L)}{\sqrt{\eta/\cosh^{2}(x/2L)+1}}
\end{equation}
$ 
\eta=\frac{b_3+2 b_{33} P_s^2}{2 b_3+b_{33} P_s^2}
$
$
L=\sqrt{\frac{g_{44}}{30 a_{111} P_s^4+2 b_3+12 b_{33} P_s^2}}
$
.
In Ref.~\cite{Behera2011} the reduced free energy $F_0$ was considered and the possibility of nonzero $P_1$ and $P_2$ was mentioned. But it will be shown that in PTO below $T_c$ the \textit{Ising} profile is in fact always stable when calculated from $F_0$. In order to get nonzero polarization at the DW center additional free energy terms are needed. 
At first, only the flexoelectric term $F_{flex}$ is considered, $f'_{14}\neq 0$, $f_{22}=0$. It implies a nonzero antisymmetric \textit{N\'eel} component $P_1\propto \frac{\partial P_3^2}{\partial x}$ (and $P_2=0$) in the whole temperature range below $T_c$ and it is in accord with the DW symmetry, $P_1(-x)=-P_1(x)$. The typical antisymmetric \textit{N\'eel} DW profile is obtained from Fig.~\ref{fig:profile} by setting $P_2=0$. For $f_{14}'>0$ it possesses head-to-head configuration (see Fig.~\ref{fig:profile}), while if $f_{14}'<0$ it has tail-to-tail configuration, it corresponds to $P_1$ in Fig.~\ref{fig:profile} taken with negative sign. For simplicity's sake we do not encounter depolarizing fields here.

\begin{figure}
	\includegraphics[width=\linewidth]{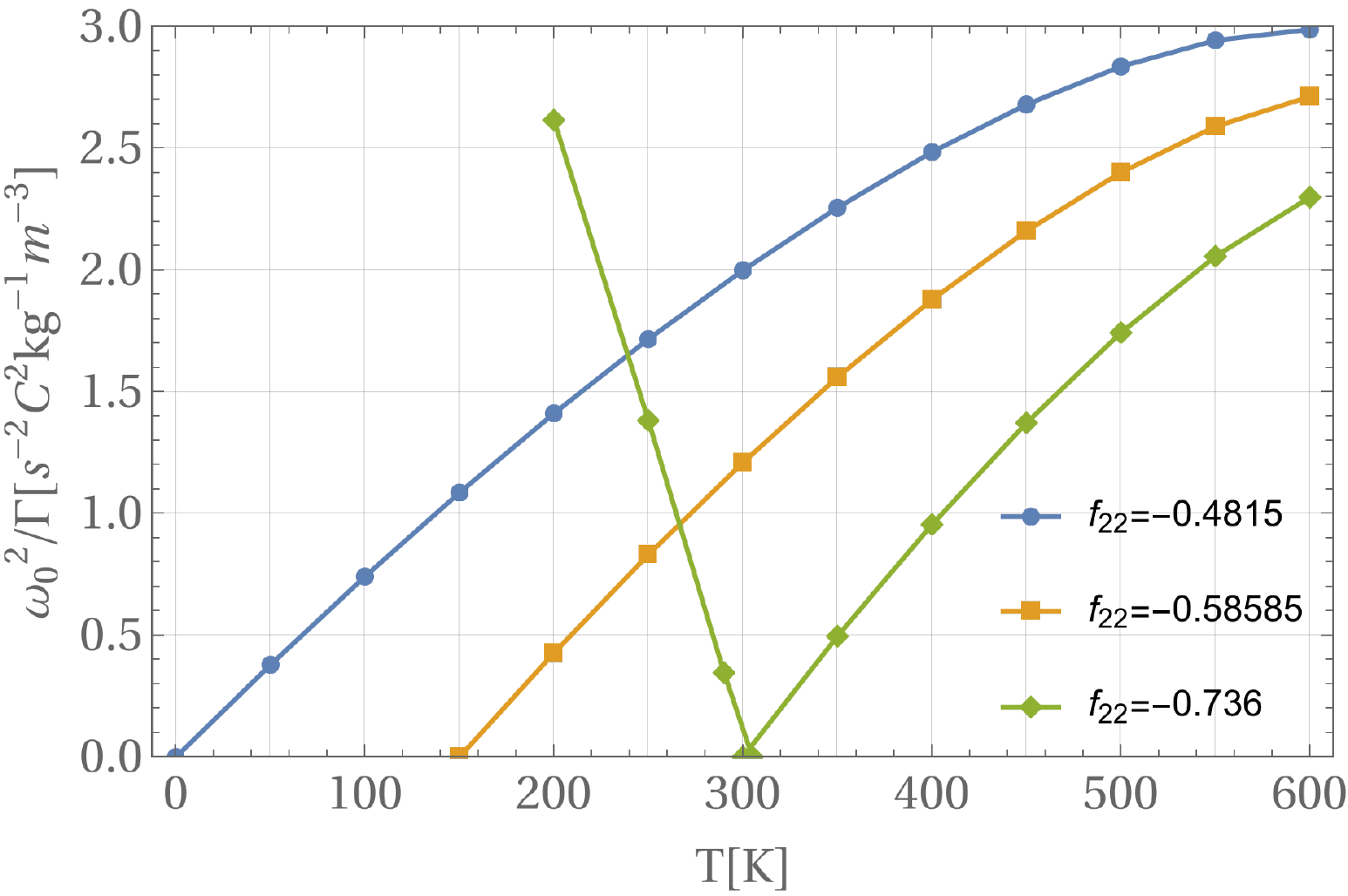}
	\caption{\label{fig:omega0}
	The temperature dependence of the DW soft mode for different values of $f_{22}$. The phase transition occurs at temperatures $T_{DW}=0,\ 150,\ 305K$. The increase of $\omega_0^2$ below $T_{DW}=305K$ is also shown.}
	\end{figure}

\begin{figure}
	\includegraphics[width=\linewidth]{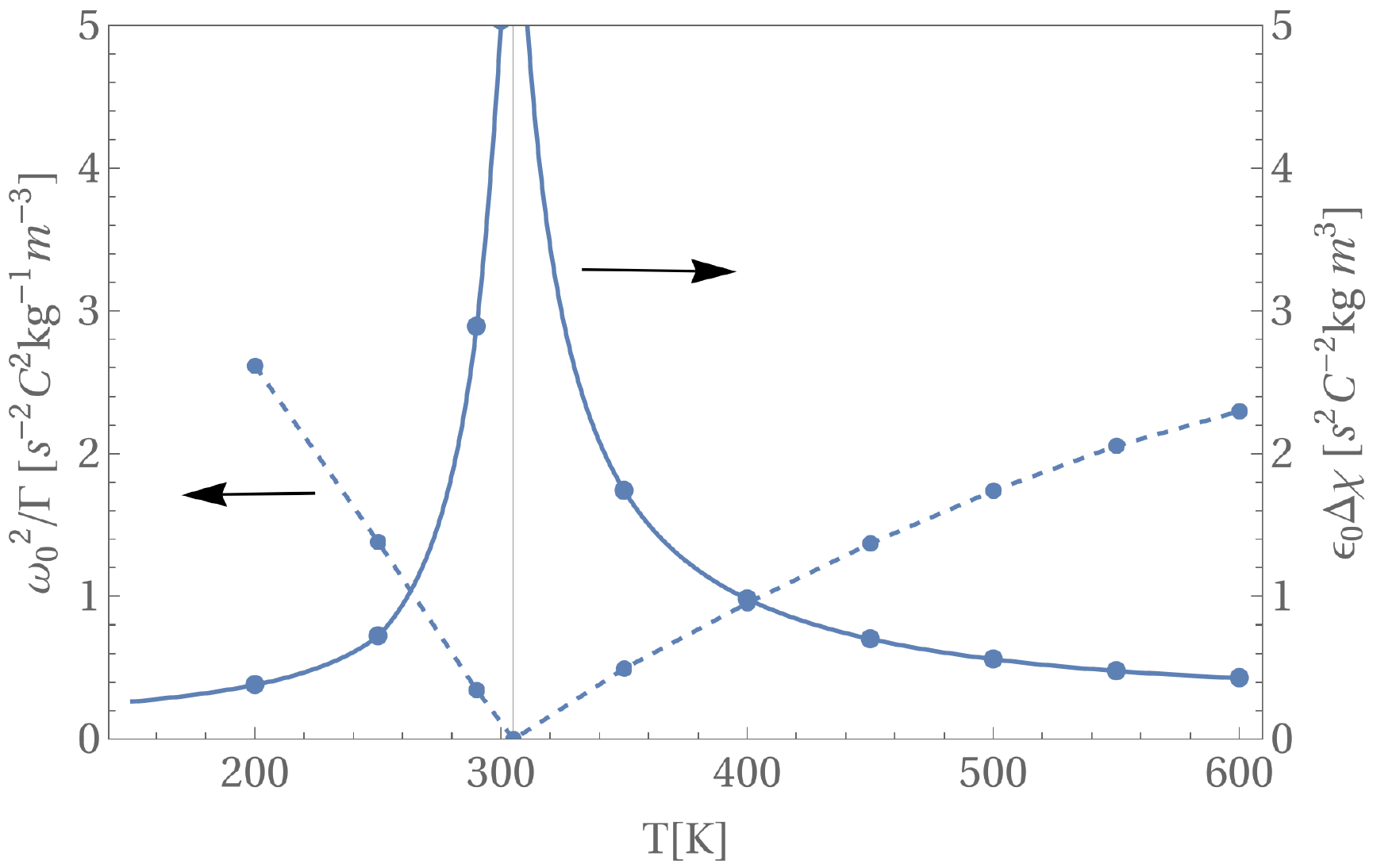}
	\caption{\label{fig:susc}
		The susceptibility divergence $\propto 1/|T-T_{DW}|$ at $T_{DW}=305K$ ($f_{22}=-0.736$). The softening of $\omega_0^2$, the same as in Fig.~\ref{fig:omega0}, is also shown for reference.}
\end{figure}

\begin{figure}
	\includegraphics[width=\linewidth]{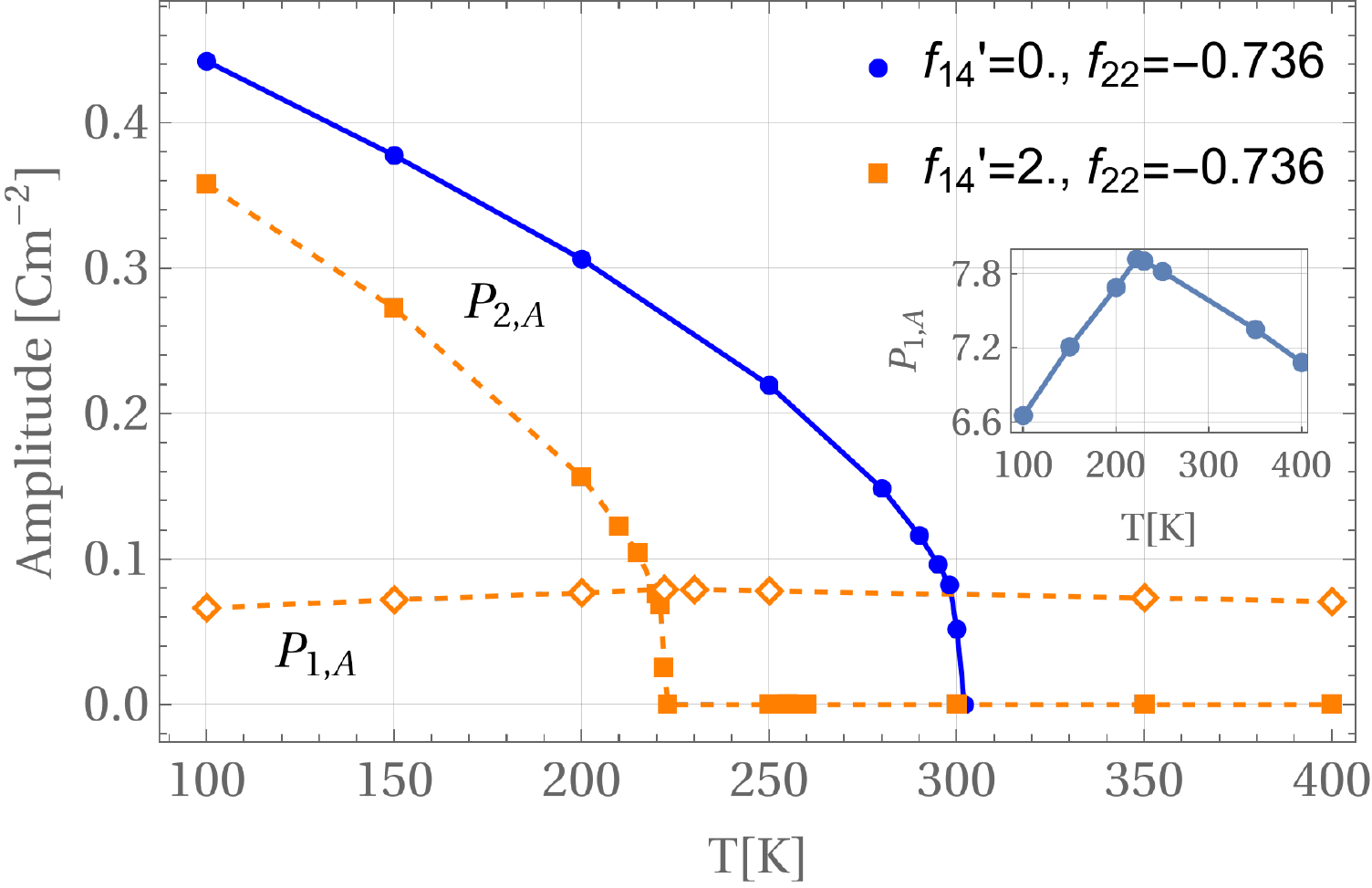}
	\caption{\label{fig:P2}
		The temperature dependence of the Bloch component $P_{2,A}$ and N\'eel component $P_{1,A}$ for 2 values of $f_{14}'$. Full line shows the \textit{Ising}$\rightarrow$\textit{Bloch} transition at $T_{DW}\approx 305K$, $P_{1,A}=0$. Dashed lines shows the transition \textit{N\'eel}$\rightarrow$\textit{Bloch}+\textit{N\'eel} at lower $T_{DW}\approx220K$, $P_{1,A}\ne0$ at all temperatures. The inset shows a tiny cusp of $P_{1,A}$ at $T_{DW}\approx220K$.} 
\end{figure}

The \textit{Bloch} DW component $P_2$ can occur by introducing a biquadratic term of the OP and its gradient. For now we assume zero flexoelectric term, $f'_{14}=0,\ f_{22}\neq 0$. The typical \textit{Bloch} profile is obtained by setting $P_1=0$ in Fig.~\ref{fig:profile}. The DW symmetry discussed above indicates that nonzero $P_2$ could appear only due to a phase transition accompanied by a decrease of the DW symmetry. Stability of the \textit{Ising} solution (\ref{eq:ising}) with respect to a small disturbance $P_2=0+\delta_2$ is inspected by solving the eigenvalue problem (equation of motion of $\delta_2$) \cite{Tagantsev2001}, see \textit{Appendix~B}:
\begin{eqnarray}\label{eq:motion}
\Gamma^{-1}\omega_0^2\delta_2= 
&-&2 \delta_2 \left(a_{112} P_3^4+b_2+b_{23} P_3^2+f_{22}
{P}_{3,x}^2\right) \\
&+&{\delta_{2,xx}} \left(2 f_{22} P_3^2+g_{44}\right)+4 f_{22} P_3
{P}_{3,x} {\delta_{2,x}} \nonumber 
\end{eqnarray}
The instability of the mode $\delta_2$ occurs when $\omega_0^2<0$. For positive $\omega_0^2$ the contribution of $\delta_2$ to the susceptibility reads $\Delta\chi=\Gamma/\varepsilon_0\omega_0^2$. $\Delta\chi$ is defined as $\Delta\chi=\delta_{2,A}/E$, where $\delta_{2,A}$ is an amplitude of the polar $\delta_{2}(x)$ mode and $E$ is an electric field, see \textit{Appendix~B}. The analytic solution of the differential equation~(\ref{eq:motion}) is unknown and we solved it numerically for several values of the biquadratic (of the OP and its gradient) coefficient $f_{22}$, Fig.~\ref{fig:omega0}. The phase transition in the DW occurs at $T_{DW}>0$ if $f_{22}<-0.4815$. The effect of negative $f_{22}$ can be seen from the quadratic $P_2^2$ term at the DW center $(b_2 + f_{22}P_{3,x}^2)P_2^2$, which decreases if $f_{22}<0$. Near above $T_{DW}$ $\omega_0^2\propto (T-T_{DW})$ (see Fig.~\ref{fig:omega0}). Below $T_{DW}$ the symmetric \textit{Bloch} component $P_2(x)$ appears, its shape is similar with $P_2(x)$ shown in Fig.~\ref{fig:profile}. Below $T_{DW}$, $\omega_0^2$ of the polar mode was calculated by solving coupled equations of motion of $\delta_2$ and $\delta_3$ obtained from (\ref{eq:A4}). It exhibits a typical hardening $\omega_0^2\propto (T_{DW}-T)$ shown in Fig.~\ref{fig:omega0}. The corresponding susceptibility $\Delta \chi$ around the phase transition at $T_{DW}=305$K exhibits a $1/|T-T_{DW}|$ divergence, Fig.~\ref{fig:susc}. 
The temperature dependence of the amplitude of the $P_2(x)$ profile is $P_{2,A}\approx (T_{DW}-T)^{1/2}$, see the solid line in Fig.~\ref{fig:P2}. A similar softening of the $P_2$ polar mode, its freeze-out below $T_{DW}$ and divergent susceptibility was obtained by the first-principles calculations in Ref.~\cite{Wojdel2014}. 

The interrelation between \textit{N\'eel} and \textit{Bloch} components comes into play when concurrently $f'_{14}\neq0$ and $f_{22}\neq 0$. The component $P_1$ exists in the whole temperature range and $T_{DW}$ is shifted to lower temperatures, see the dashed lines in Fig.~\ref{fig:P2}. Below $T_{DW}$ the $P_1$ and $P_2$ components coexist. The inset in Fig.~\ref{fig:P2} shows that $P_1$ exhibits a tiny anomaly at $T_{DW}$. The polar mode softening and the anomalous susceptibility are similar as shown in Fig.~\ref{fig:susc} for the previous case. 

\begin{table*}[t] 
	\caption{Free energy parameters \cite{Haun1987},\cite{Li2002},\cite{Behera2011}.\label{tab:1}}
		\begin{tabular}{c r | r r | r l}
			\hline\hline
			$\alpha_{1}$ & $3.8(T-752 \mathrm{K})*10^5\mathrm{C}^{-2}\mathrm{m}^2\mathrm{N}$ 
			&$Q_{11}$ & $0.089\mathrm{C}^{-2}\mathrm{m}^{4}$
			&$g_{11}$ & $2.0*10^{-10}\mathrm{m}^{4}\mathrm{C^{-2}N}$ \\
			$\alpha_{11}$ & $-0.73*10^8\mathrm{C}^{-4}\mathrm{m}^6\mathrm{N}$
			&$Q_{12}$ & $-0.026\mathrm{C}^{-2}\mathrm{m}^{4}$
			&$g_{44}$ & $1.0*10^{-10}\mathrm{m}^{4}\mathrm{C^{-2}N}$ \\
			$\alpha_{12}$ & $7.5*10^8\mathrm{C}^{-4}\mathrm{m}^6\mathrm{N}$
			&$Q_{44}$ & $0.0337\mathrm{C}^{-2}\mathrm{m}^{4}$
			\\
			$\alpha_{111}$ & $2.6*10^8\mathrm{C}^{-6}\mathrm{m}^{10}\mathrm{N}$
			&$s_{11}$ & $8.0*10^{-12}\mathrm{m}^{2}\mathrm{N}^{-1}$ \\
			$\alpha_{112}$ & $6.1*10^8\mathrm{C}^{-6}\mathrm{m}^{10}\mathrm{N}$
			&$s_{12}$ & $-2.5*10^{-12}\mathrm{m}^{2}\mathrm{N}^{-1}$ \\
			$\alpha_{123}$ & $-37*10^8\mathrm{C}^{-6}\mathrm{m}^{10}\mathrm{N}$
			&$s_{44}$ & $9.0*10^{-12}\mathrm{m}^{2}\mathrm{N}^{-1}$ \\
		\end{tabular}
\end{table*}

\textit{Summary}---
The symmetry of $180^\circ$-DW indicates the existence of unswitchable antisymmetric \textit{N\'eel} $P_1$ polarization at the DW center in the whole temperature range below $T_c$ and within the Landau-Ginzburg description it is indeed induced by the flexoelectric term. The depolarizing charges should diminish $P_1$, but for simplicity they are not considered here. Similar results concerning the flexoelectric term were also obtained by phase-field modeling \cite{Wang2017}. Here we have shown, that the symmetric switchable \textit{Bloch} polarization $P_2$ occurs due to a phase transition in the domain wall at $T_{DW}$, which is driven by the biquadratic coupling of the OP and its gradient. The softening of the polar mode, divergent susceptibility and the temperature dependence of $P_2$ below $T_{DW}$ are in excellent agreement with the results from  first-principles calculations  \cite{Wojdel2014,Iniguez2020}. 

This work was supported by Operational Program Research, Development and Education (financed by European Structural and Investment Funds and by the Czech Ministry of Education, Youth, and Sports), Project No. SOLID21-CZ.02.1.01/0.0/0.0/16\_019/0000760). 
%
\\[5mm]
\setcounter{equation}{0}
\renewcommand{\theequation}{A.\arabic{equation}}
\textit{Appendix~A}:\\
	The $'b'$ coefficients in $F_0$:
	\begin{eqnarray}
		b_1&=& a_1-\frac{P_s^2 Q_{12} (Q_{11}+Q_{12})}{s_{11}+s_{12}},\nonumber\\
		b_2&=& a_1+\frac{P_s^2 \left(s_{12} \left(Q_{11}^2+Q_{12}^2\right)-2 Q_{11} Q_{12}
			s_{11}\right)}{s_{11}^2-s_{12}^2},\nonumber\\
		b_3&=& a_1-\frac{P_s^2 \left(s_{11} \left(Q_{11}^2+Q_{12}^2\right)-2 Q_{11} Q_{12} s_{12}\right)}{s_{11}^2-s_{12}^2}, \nonumber\\
		b_{11}&=& a_{11}+\frac{Q_{12}^2}{s_{11}+s_{12}}, \\
		b_{12}&=& a_{12}+\frac{Q_{12} (Q_{11}+Q_{12})}{s_{11}+s_{12}},\nonumber\\
		b_{22}&=& a_{11}+\frac{s_{11} \left(Q_{11}^2+Q_{12}^2\right)-2 Q_{11} Q_{12} s_{12}}{2 s_{11}^2-2 s_{12}^2},\nonumber\\
		b_{23}&=& a_{12}-\frac{s_{12}
			\left(Q_{11}^2+Q_{12}^2\right)-2 Q_{11} Q_{12} s_{11}}{s_{11}^2-s_{12}^2}+\frac{Q_{44}^2}{2
			s_{44}} \nonumber
	\end{eqnarray}
\\
\setcounter{equation}{0}
\renewcommand{\theequation}{B.\arabic{equation}}
\textit{Appendix~B}:\\
The free energy functional and its variation,
\begin{equation}
	\mathcal{L}=\int_{-\infty}^{\infty}F(\mathbf{P}(x),\partial_x\mathbf{P}(x))dx
\end{equation}
\begin{equation}
\delta\mathcal{L}=\int_{-\infty}^{\infty}\frac{\delta \mathcal{L}}{\delta\mathbf{P}}\delta\mathbf{P}dx=\int_{-\infty}^{\infty}\left(\frac{\partial L}{\partial\mathbf{P}}-\frac{d}{dx}\frac{\partial L}{\partial \mathbf{\dot{P}}}\right)\delta\mathbf{P}dx
\end{equation}
The DW profiles $\boldsymbol{\mathcal{P}}$ are solutions of 3 equilibrium equations:
\begin{equation}
\frac{\delta \mathcal{L}}{\delta P_i}\equiv
\left(\frac{\partial L}{\partial P_i}-\frac{d}{dx}\frac{\partial L}{\partial \dot{P_i}}\right) =0,\ i=1,2,3
\end{equation} 
A small perturbation  
$\mathbf{P}=\boldsymbol{\mathcal{P}}+\boldsymbol{\delta}$
leads to 3 equations of motion:
\begin{equation}\label{eq:A4}
\Gamma^{-1}\ddot{\delta_i}=-\Gamma^{-1}\omega_0^2 \delta_i =
\left.-\frac{\delta \mathcal{L}}{\delta P_i}\right\vert_{\mathbf{P}\rightarrow\boldsymbol{\mathcal{P}}+\boldsymbol{\delta}}
\end{equation} 
where in the right-hand side only the linear terms in $\boldsymbol{\delta}$ are kept. The perturbation is assumed as  $\boldsymbol{\delta}\propto e^{i\omega x}$, the coefficient $\Gamma=ne^2/m$, where $m,\ e,\ n$ are mass, charge and density of ions, respectively, $[\Gamma]=kg^{-1}m^{-3}C^{2}$. The DW profile is stable when the smallest eigenvalue $\omega_0^2>0$. The static susceptibility of the polar eigenmode $\delta_2(x)$ is defined as  $\Delta\chi \equiv \delta_{2,A}/E=\Gamma/\varepsilon_0\omega_0^2$, where $\delta_{2,A}$ is an amplitude of the polar mode. In case of the \textit{Ising} profile 3 equations (\ref{eq:A4}) are decoupled.

\bibliography{pto_paper.bib}

\end{document}